\documentclass[journal,twoside,web]{ieeecolor}
\usepackage{generic}
\usepackage{cite}
\usepackage{amsmath,amssymb,amsfonts}
\usepackage{algorithmic}
\usepackage{graphicx}
\usepackage{textcomp}
\def\BibTeX{{\rm B\kern-.05em{\sc i\kern-.025em b}\kern-.08em
    T\kern-.1667em\lower.7ex\hbox{E}\kern-.125emX}}
\begin{document}
\title{Convolutional Neural Networks Utilizing Multifunctional Spin-Hall MTJ Neurons}
\author{Andrew W. Stephan and Steven J. Koester, \IEEEmembership{Fellow, IEEE}
\thanks{Manuscript submitted \today. This work was supported by Seagate Technology PLC.}
\thanks{The authors acknowledge the Minnesota Supercomputing Institute (MSI) at the University of Minnesota for providing resources that contributed to the research results reported within this paper. URL: http://www.msi.umn.edu}
\thanks{A. W. Stephan is with the College of Science and Engineering, University of Minnesota, Minneapolis, MN 55455 USA (e-mail:steph506@umn.edu).}
\thanks{S. J. Koester is with the College of Science and Engineering, University of Minnesota, Minneapolis, MN 55455 USA (e-mail:skoester@umn.edu).}}

\maketitle

\begin{abstract} 
We propose a new network architecture for standard spin-Hall magnetic tunnel junction-based spintronic neurons that allows them to compute multiple critical convolutional neural network functionalities simultaneously and in parallel, saving space and time. An approximation to the Rectified Linear Unit transfer function and the local pooling function are computed simultaneously with the convolution operation itself. A proof-of-concept simulation is performed on the MNIST dataset, achieving up to 98\% accuracy at a cost of less than 1 nJ for all convolution, activation and pooling operations combined. The simulations are remarkably robust to thermal noise, performing well even with very small magnetic layers.
\end{abstract}


\section{Introduction}
\label{sec:introduction}
The magnetic tunnel junction (MTJ) is ubiquitous in spintronics, with applications in memory, logic and neuromorphic computing\cite{KRoy,Naeemi1,Naeemi2,mLogic,Spin Switch}. One of the most common spintronic neuron implementations relies on a voltage divider composed of two MTJs and a heavy metal layer to switch one of the free layers (FLs) via spin torque induced by the spin Hall effect (SHE). The net tunneling resistance of the structure depends on the FL orientation relative to that of the pinned layer (PL). This structure is quite versatile. Depending on the geometry of the structure, the possibilities include a simple step threshold neuron, a non-step neuron, an integrate-and-spike neuron or a stochastic spike neuron\cite{KRoy, Naeemi1, Naeemi2}. One of the more prolific models of neuromorphic computing is the convolutional neural network (CNN) which is particularly useful for image classification\cite{CoNNs}. We propose a novel extension of the SHE-MTJ neurons to the CNN framework which will greatly reduce the time, energy and number of devices required to process images by parallelizing the CNN functions. A simulation of this network promises remarkable image classification accuracy even if the elements involved are far too thermally unstable for digital logic.

\section{Design}
\label{sec:Design}
The basic neuron structure consists of a pair of MTJs which act as a voltage divider. A high resistance and large tunneling magnetoresistance ratio (TMR) is preferred to reduce leakage and increase output range\cite{Giant TMR}. The output voltage is read by an inverter pair operating in the linear regime. One of the MTJ free layers (FLs) is joined to a heavy metal that injects spin current into the FL based on the charge current injected into it by a set of inverters. Weights are applied by resistors between each inverter and the heavy metal contact (see Fig. \ref{fig:SingleDevice}). The precise type of resistor or memristor used is not treated here. If the input impedance is sufficiently low compared to the resistance of the weights the inverters act as parallel current sources. The current reaching the heavy metal thus represents a multiply-accumulate (MA) operation\cite{KRoy, Naeemi1, Naeemi2, Crossbar1, Crossbar2}. The cell output voltage as a function of FL magnetization is treated by a 1st-order approximation based on an HSPICE simulation using the 16-nm technology node transistor model as in\cite{Jiaxi,PTM}. 

\subsection{Multiply-Accumulate-Activate Sequence}
\begin{figure}
\centering
\includegraphics[scale=.5]{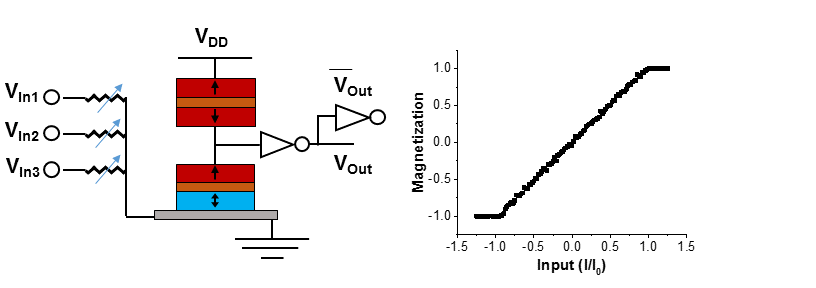}
\caption{Device structure and magnetic response of FL driven along its hard axis by SHE. The response is a very close fit to identity between $\pm$ 1 and saturates thereafter. Input is defined by the critical current $I_0$ required to saturate the FL.}
\label{fig:SingleDevice}
\end{figure}

\begin{figure}
\centering
\includegraphics[scale=.45]{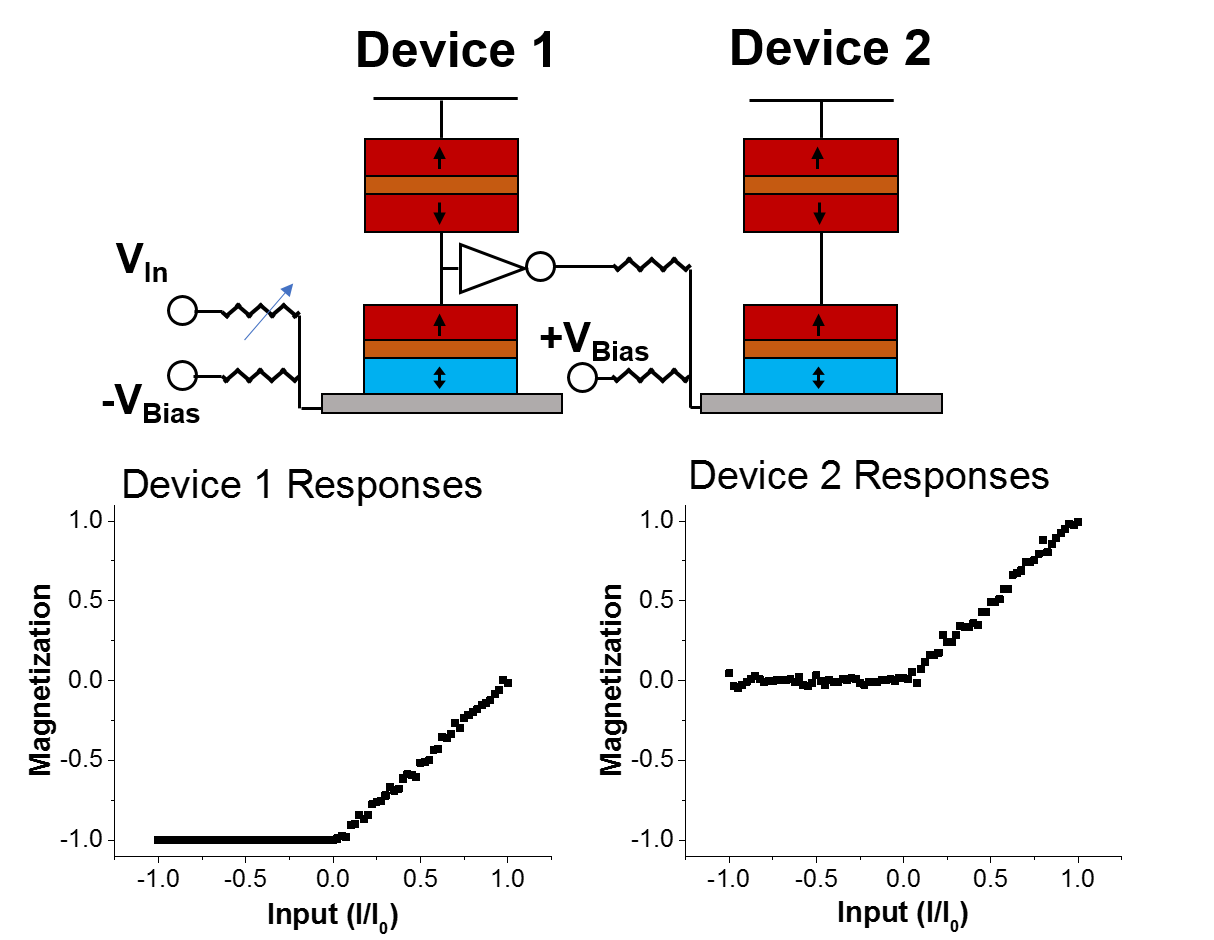}
\caption{Magnetic response of an MAA sequence. Each device has the same response characteristic as in Fig. \ref{fig:SingleDevice}. Device 1 has a $-1$ input bias while Device 2 has a $+1$ input bias. The result is a close approximation to the ReLU function for inputs less than $+1$. Input is defined by the critical current $I_0$ required to saturate the FL.}
\label{fig:ReLU}
\end{figure}

The Rectified Linear Unit (ReLU) is ubiquitous as an activation function in software-based neural networks for its superior results and ease of use in training weights. We propose a novel arrangement of spintronic elements that will approximate the ReLU activation. It was pointed out by Lou \emph{et al.}\cite{CeNNs with CoNNs} that a device with a saturated linear transfer function can be used as the basis for a simple in-hardware implementation of the ReLU. To that end, we use a simple variation on the usual SHE-MTJ neuron. By using in-plane anisotropy but assuming perpendicular SHE torque\cite{PMA1, PMA2} to direct the torque along the FL hard axis, we produce a neuron with a saturated linear transfer function (see Fig. \ref{fig:SingleDevice}). Two of these devices in sequence with $-1$ and $+1$ bias added to their respective inputs produces an output that closely resembles the ReLU so long as the initial net input, excluding bias, is at or below the level required to saturate the magnet in the positive direction\cite{CeNNs with CoNNs} (see Fig. \ref{fig:ReLU}). The input is mapped to an actual current value as will be described in Sec. \ref{sec:Simulation}. The simulation used to predict this behavior is also described therein.

\subsection{Pooling}
\begin{figure*}
\centering
\includegraphics[scale=.47]{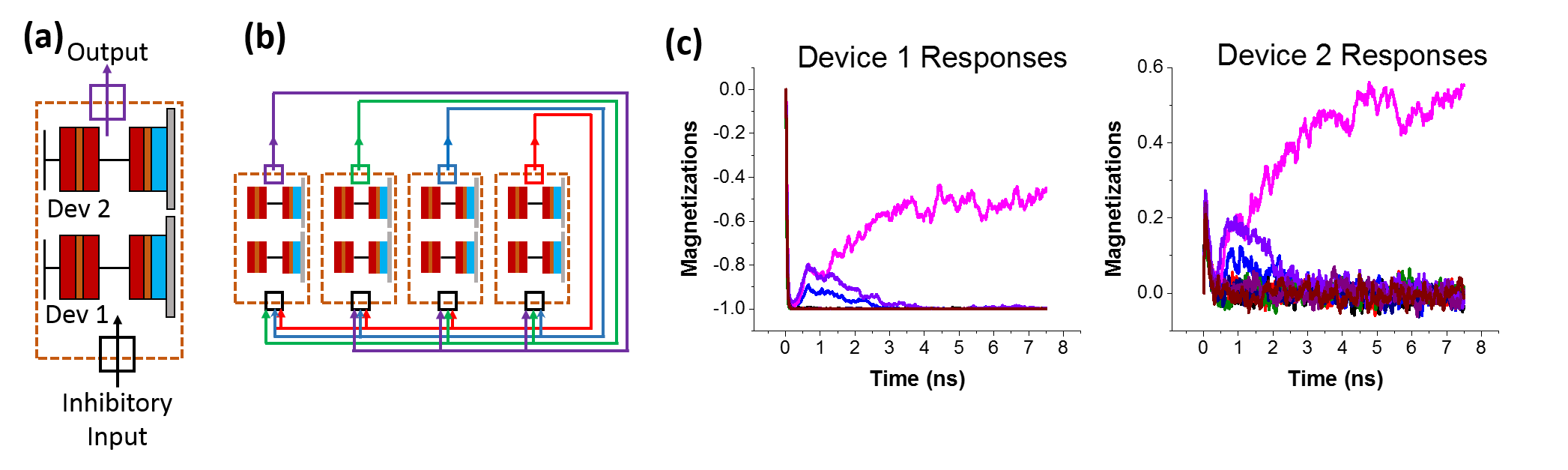}
\caption{ (a) Circuit symbol for MAA sequence with output and an inhibitory input indicated. The external input and biases are not shown for simplicity. (b) A MAAP set comprising four MAA sequences. Each MAA sequence receives an inhibitory input from its three companions, resulting in a winner-take-all competition. (c) Magnetic response of an MAAP set comprising nine MAA sequences. The first and second neurons of each sequence are both displayed. All but the sequence with the largest external input have their output neurons driven to the zero state. The MAAP output is read by summing the result of all MAA sequences.}
\label{fig:BigMAAP}
\end{figure*}

\begin{figure}
\centering
\includegraphics[scale=.45]{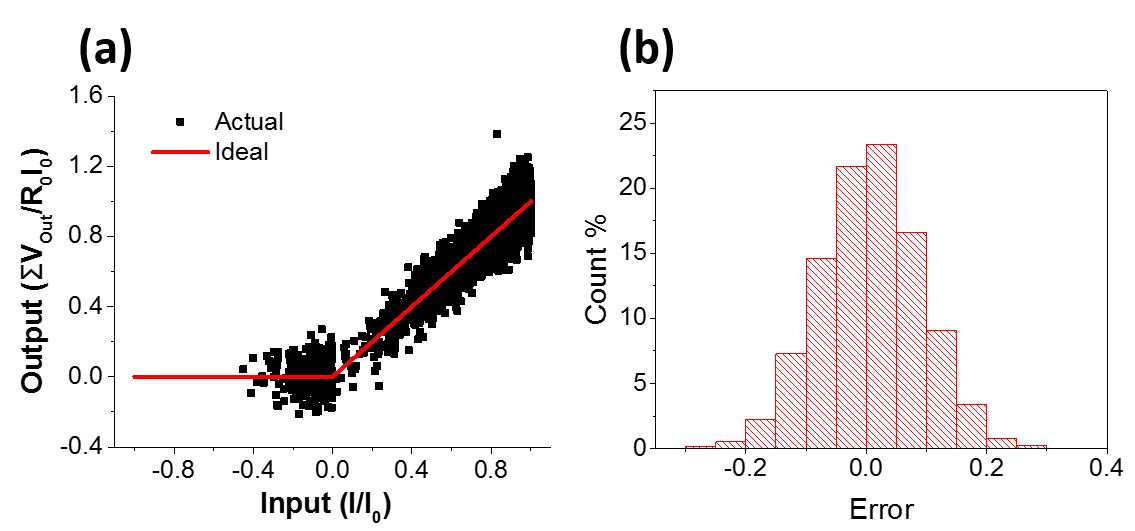}
\caption{Result of a 4000-iteration Monte Carlo simulation of a nine-input MAAP set to determine output error. Input and output are defined in terms of the critical current $I_0$ required to saturate the FL and the unit weight resistance $R_0$. (a) The actual results (black squares) vs. the input, plotted alongside the ideal ReLU result (red line). (b) Histogram of the errors from (a).}
\label{fig:MAAPHist}
\end{figure}

In addition to an activation function such as the ReLU, one other capability is required to build a CNN. A local pooling function, which takes the maximum of a set of MAA outputs, is necessary. Achieving this requires only a simple change, with no additional neurons and only a few CMOS elements. At each of the set of outputs to be pooled, a negatively-weighted output connection is added which joins to the input of every other MAA sequence in the set. The layout described here can be seen for a simplified four-MAA sequence set in Fig. \ref{fig:BigMAAP} (b). The MAA sequences compete in a winner-take-all contest which ends when the sequence with the largest overall input drives the inputs of the other sequences to zero, thus eliminating the competition, and finally sets its own output to the value demanded by its external inputs (see Fig. \ref{fig:BigMAAP}). We note that the smaller the difference between the maximum input and the rest, the longer this process takes to separate them. The output is collected from all members of the set and summed, giving the maximum without needing to know which produced it. This simultaneous multiply-accumulate-activate-pool (MAAP) set was tested with a Monte Carlo simulation to determine the accuracy of its output. The physical parameters of the simulation are given in Table \ref{tab:parameters}. Each simulation consisted of a MAAP set comprising 9 MAA sequences, each with a random external input between $\pm$ 1. The final normalized result is determined by summing all nine output voltages and dividing by $R_0I_0$, where $R_0$ is the unit resistance value as described in Sec. \ref{sec:Simulation}. This output should be equal to the rectified value of the largest of the nine external normalized inputs. The results are shown in Fig. \ref{fig:MAAPHist}. The red line indicates an ideal ReLU activation function while the cluster of black squares indicates the actual output of the simulated MAAP sets. The error resembles a Gaussian random variable with mean approximately equal to the ideal value. The width of the peak varies depending on the level of thermal noise.

\section{Simulation}
\label{sec:Simulation}

\begin{table}
\centering
\caption{Simulation Parameters}
\label{table}
\setlength{\tabcolsep}{3pt}
\begin{tabular}{|p{25pt}|p{100pt}|p{75pt}|}
\hline
Symbol& 
Quantity& 
Value \\
\hline
\vspace{0.005in}
$K$& 
\vspace{0.005in}
crystalline anisotropy& 
\vspace{0.005in}
$4.5$ kJ/m$^3$ \\
$V$& 
ferromagnet volume& 
50 - 800 nm$^3$ \\
$L$ &
ferromagnet side length  &
5 - 20 nm \\
$\Delta$ &
thermal stability factor &
4 - 60 \\
$M_S$& 
saturation magnetization& 
0.5 MA/m \cite{Heusler1, Heusler2, Heusler3}\\
$\alpha$&
Gilbert damping&
0.01\\
$t_{HM}$ &
heavy metal thickness &
5 nm \\
$\theta$ &
spin-Hall angle &
0.3\cite{Tungsten,Tantalum} \\
$V_{DD}$ &
sensing voltage&
500 mV \\
$R_0$ &
unit weight resistance&
6.4-13.5 k$\Omega$ \\
$I_{c0}$& 
unit input current & 
37-78 $\mu$A \\
$R_M$ &
MTJ high/low resistance &
250 k$\Omega$ / 2.5 M$\Omega$ \\
$R_R$ &
reference MTJ resistance &
800 k$\Omega$ \\
$\Delta$t &
simulation time step &
10 ps \\
$V_T$ &
transistor threshold voltage &
0.2 V \\
\hline
\multicolumn{3}{p{200pt}}{}\\
\end{tabular}
\label{tab:parameters}
\end{table}

The FL is modeled using the macrospin approximation. The Landau-Lifshitz-Gilbert equation relates the motion of the unit magnetization $\boldsymbol{m}$ to the net field
\begin{gather}
\frac{d\boldsymbol{m}}{dt} = -\gamma \mu_0 \Big( (\boldsymbol{m} \times \boldsymbol{H_{Eff}}) - \alpha \big( \boldsymbol{m} \times (\boldsymbol{m} \times \boldsymbol{H_{Eff}}) \big) \Big),
\end{gather}
where $\gamma$ is the gyromagnetic ratio, $\mu_0$ is the vacuum permeability, $\alpha$ is the Gilbert damping and $\boldsymbol{H_{Eff}}$ is the effective field. This term contains the intrinsic thermal, anisotropy and demagnetization fields as well as the external SHE-induced field $\boldsymbol{H_{SHE}}$. The thermal field is given by a multivariate Gaussian random variable with zero mean and variance matrix
\begin{gather}
\boldsymbol{\Sigma} = \boldsymbol{I} \sqrt{\frac{2k_BT\alpha}{\gamma M_SV\Delta t}},
\label{HT}
\end{gather}
where $\boldsymbol{I}$ is the identity matrix and $k_B$, $T$, $V$ and $\Delta t$ are the Boltzmann constant, temperature, magnetic volume and simulation time step respectively. The crystalline anisotropy field is given by 
\begin{gather}
\boldsymbol{H_K} = 2\frac{\boldsymbol{K}\cdot \boldsymbol{m}}{M_S}
\label{HK}
\end{gather}
where $\boldsymbol{K}$ is the anisotropy matrix. The demagnetization field is estimated using the approximation
\begin{gather}
\boldsymbol{H_D} = -M_S\frac{\{l_{FM}^{-2}m_x, w_{FM}^{-2}m_y, t_{FM}^{-2}m_z \}}{l_{FM}^{-2} + w_{FM}^{-2} + t_{FM}^{-2}}
\label{HD}
\end{gather}
where $l_{FM}$, $w_{FM}$ and $t_{FM}$ are the dimensions of the FL and $m_x$ = $\boldsymbol{m}\cdot \boldsymbol{\hat{x}}$, etc. The SHE-field is approximated by\cite{Ralph and Stiles, Ispin}
\begin{gather}
\boldsymbol{H_{SHE}} = \frac{1}{\mu_0} \frac{I_S}{2q} \frac{\hbar}{\alpha V M_S}\boldsymbol{\hat{z}},
\label{HSHE}
\end{gather}
where $I_S$ is the magnitude of the spin current and $V$ is the volume of the magnet. The spin current is proportional to the charge current $I_C$
\begin{gather}
I_S = \theta \frac{l_{FM}}{t_{HM}}I_C,
\label{ISPIN}
\end{gather}
where $\theta$ is the spin Hall angle and $t_{HM}$ is the thickness of the heavy metal. The various physical parameters for the simulation are found in Table \ref{tab:parameters}. Magnetic values consistent with Heusler alloys were chosen\cite{Heusler1, Heusler2, Heusler3}. The power consumption in a MAAP set is dominated by the $I^2 R$ loss caused by the input currents passing through their respective weight resistors. Further discussion of power consumption is given in Sec. \ref{sec:Results}. 

\begin{figure*}
\centering
\includegraphics[scale=0.45]{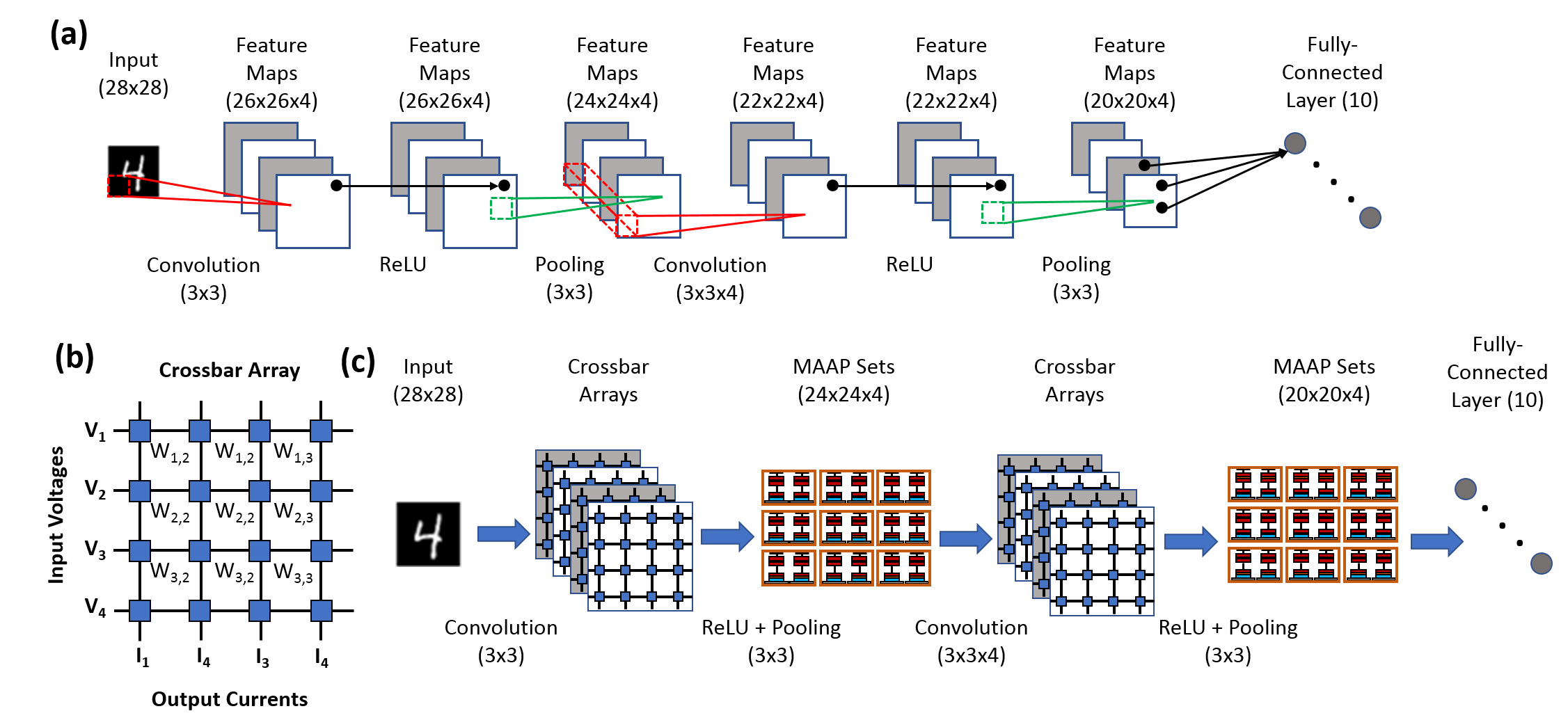}
\caption{(a) Convolutional Neural Network used to train weights for MNIST handwritten digit image classification. (b) Crossbar array structure used to perform convolution. (c) MAAP architecture version of CNN from (a). The ReLU and Pooling layer pairs are combined into a single layer of MAAP sets into which the crossbar array output currents are fed.}
\label{fig:MAAPCNN}
\end{figure*}

\begin{figure}
\centering 
\includegraphics[scale=0.45]{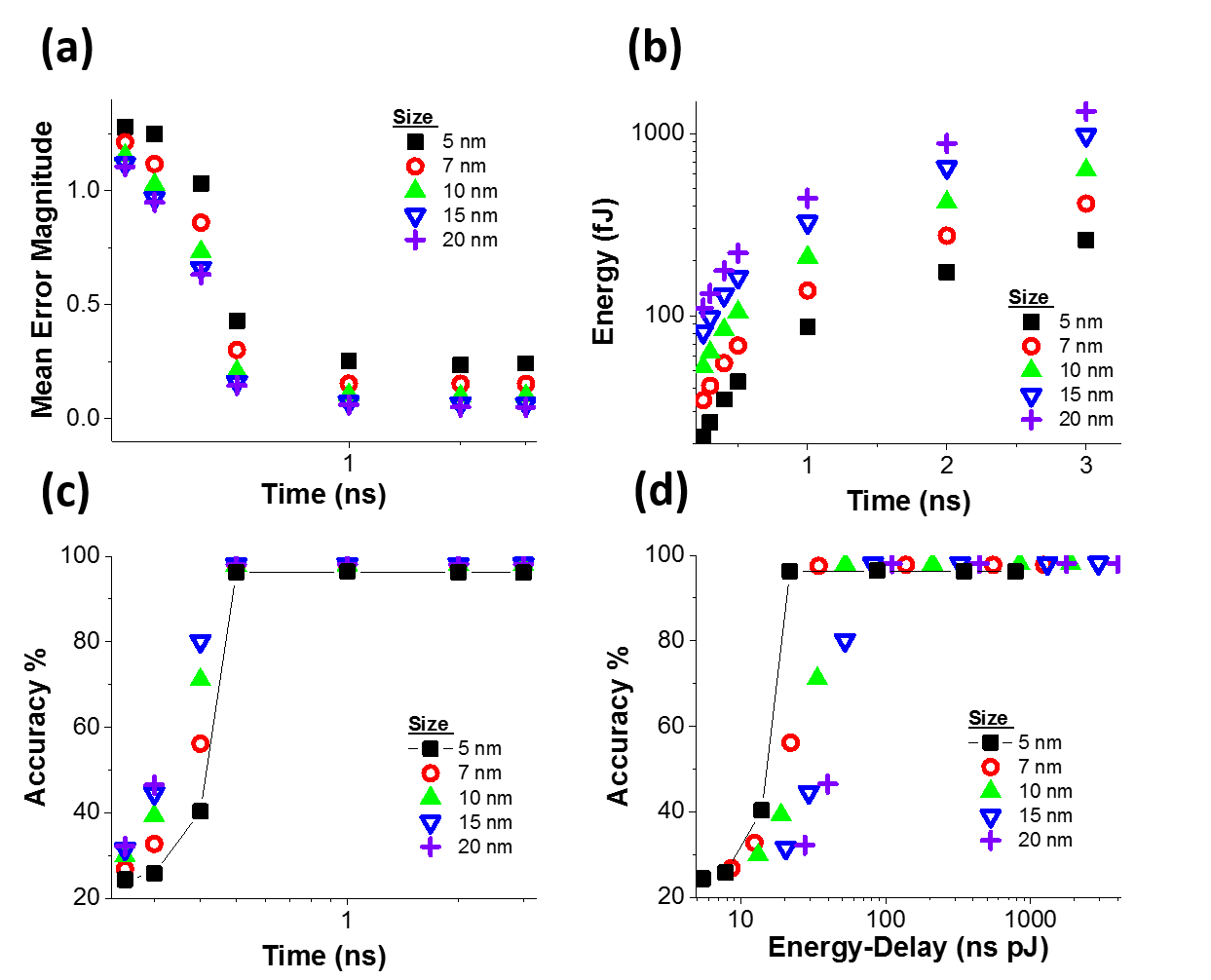}
\caption{(a) Average magnitude of normalized MAAP set output error vs. operational runtime with MTJ square side length as a parameter. Decreasing the size increases thermal noise and thus error. (b) Energy consumed per MAAP set vs. operational runtime of each MAAP set. Decreasing the size reduces the critical charge current, saving energy. (c) Image classification accuracy of MAAP-CNN vs. operational runtime. (d) Classification accuracy vs. energy-delay product of MAAP sets.}
\label{fig:ED}
\end{figure}

\begin{figure}
\centering 
\includegraphics[scale=0.45]{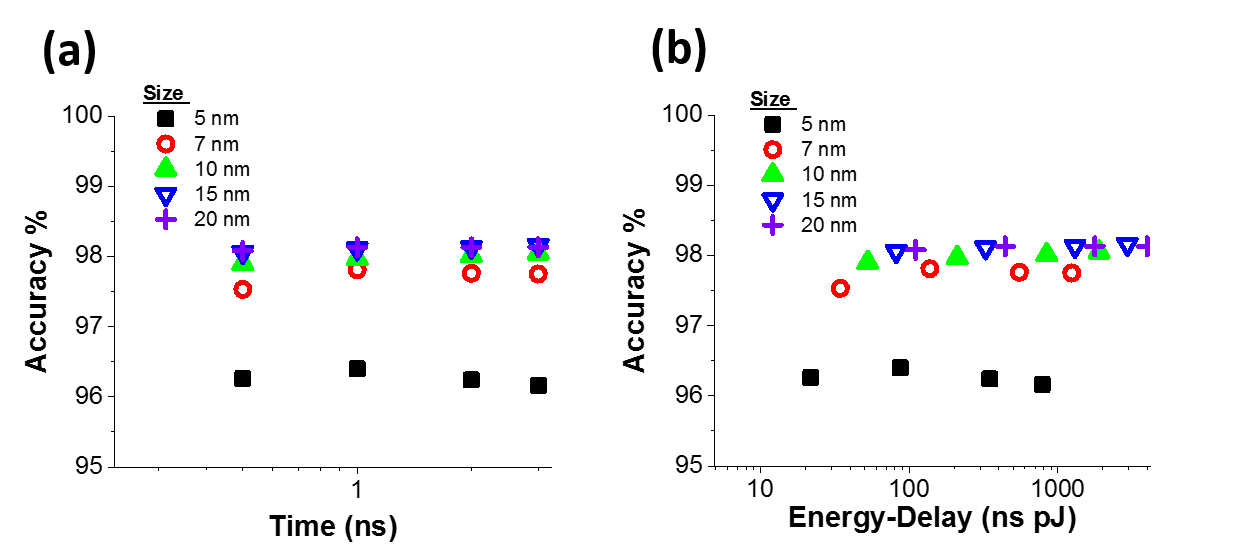}
\caption{(a) Close-up of maximum image classification accuracy of MAAP-CNN vs. operational runtime. (b) Close-up of maximum image classification accuracy vs. energy-delay product of MAAP sets.}
\label{fig:EDTop}
\end{figure}

In order to map a neural network to these SHE-based neurons, we define the critical charge current $I_{c0}$:
\begin{gather}
I_{c0} = 2 \mu_0 \frac{q}{\hbar} \frac{\alpha}{\theta} (4 \pi N_Z M_S^2) (\frac{t_{HM}}{l_{FM}}),
\label{Ic0}
\end{gather}
where $N_Z$ is the third component of the demagnetization tensor (\ref{HD}). A charge current of this magnitude will suffice to saturate the FL of a neuron, so $I_{c0}$ is the unit current corresponding to a CNN input of `1'. In general the mapping of a CNN input value $x$ to the SHE neurons is
\begin{gather}
I_{In} = x \cdot I_{c0}.
\label{Iin}
\end{gather}
The maximum output of the inverters used in the network is 0.5 V. This maximum output should produce a unit current, and so we define the unit weight resistance $R_0$:
\begin{gather}
R_0 = \frac{0.5}{I_{c0}}.
\label{R0}
\end{gather}
A weight of $w$ is encoded by $R = R_0/w$.
\section{Results}
\label{sec:Results}
\subsection{Ideal Devices}

The MAAP structure was tested using the MNIST handwritten digits dataset. In order to obtain the weights, a network was written and trained using the TensorFlow API. The structure of this network is shown in Fig. \ref{fig:MAAPCNN} (a). All convolutional and pooling kernels were  3 x 3 with a stride of 1. This standard CNN was then mapped to a MAAP-CNN architecture as seen in Fig. \ref{fig:MAAPCNN} (b)-(c). Each pooling layer with its preceding convolution and ReLU activation layers became a collection of MAAP sets and crossbar arrays, which were simulated processing the MNIST data. To preserve computational resources the outcomes of the MAAP sets were predicted by a random variable based on the previously collected error data(see Fig. \ref{fig:MAAPHist}). This was done for five different versions of the MAAP architecture using MTJ FLs with in-plane dimensions of 5x5, 7x7, 10x10, 15x15 and 20x20 nm$^2$ with a constant thickness. The normalized MAAP output error is plotted vs. MAAP runtime in Fig. \ref{fig:ED} (a). We note that the MAAP output error levels off at about 1 ns of runtime. As the MTJ area increases, the greater thermal stability results in lower error. The energy usage per MAAP set is plotted against runtime in Fig. \ref{fig:ED} (b). A typical positive correlation is shown, with larger MTJs requiring greater energy expenditure. The MAAP-CNN image classification accuracy is plotted against runtime in Fig. \ref{fig:ED} (c) showing that the CNN is quite robust to MAAP output errors even down to 0.5 ns runtime. Although the accuracy is roughly constant for any runtime greater than 0.5 ns, the highest achievable accuracy is dictated by the MTJ size as shown in Fig. \ref{fig:EDTop} (a). Finally, the classification accuracy is plotted against the MAAP set energy-delay product in Fig. \ref{fig:ED} (d). Again the relationship between maximum achievable accuracy, energy-delay and MTJ size is displayed in greater detail in Fig. \ref{fig:EDTop} (b). 

\subsection{Device Variation}

\begin{figure}
\centering
\includegraphics[scale=0.45]{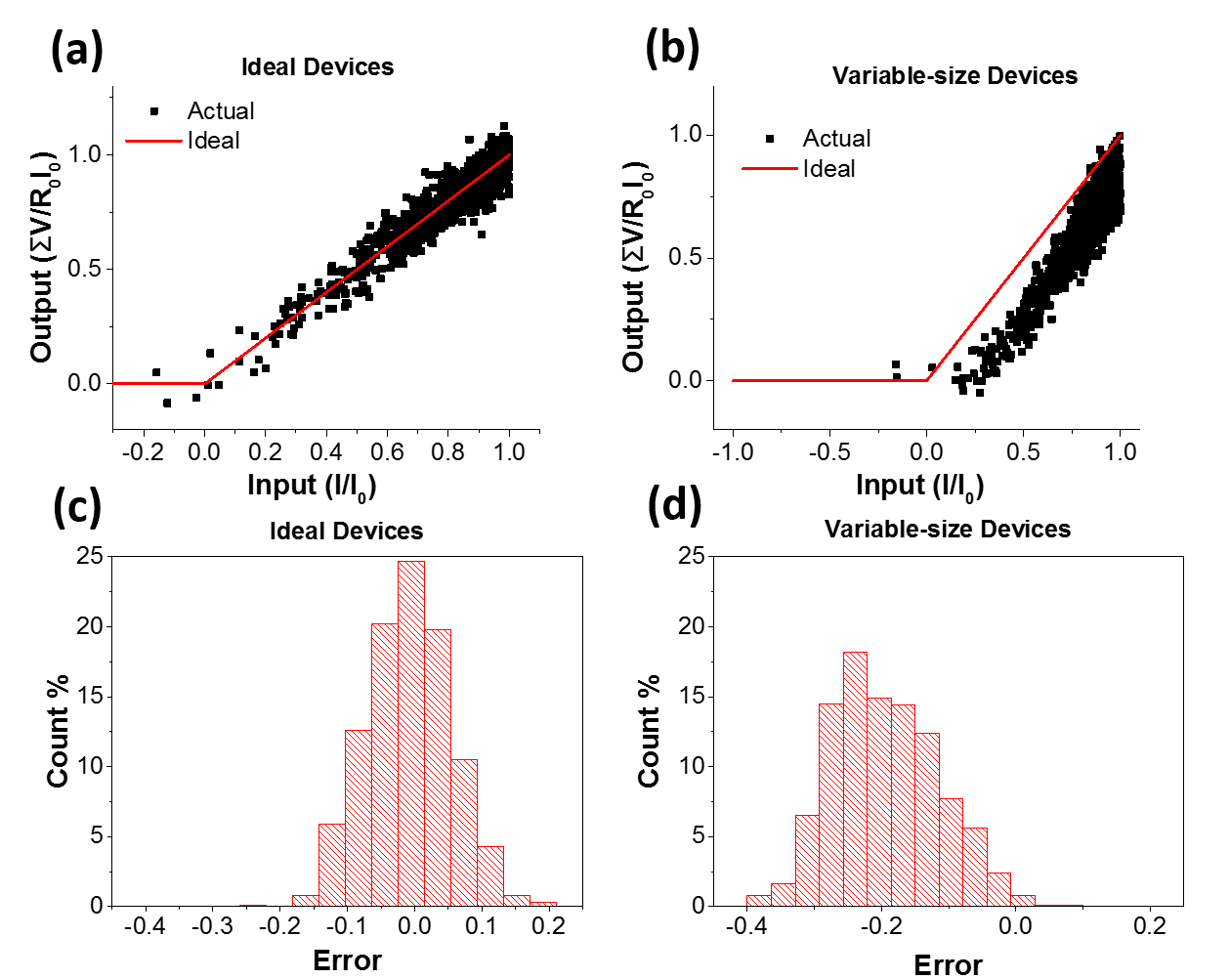}
\caption{(a)-(b) Result of a 1000-iteration Monte Carlo simulation to determine output error of nine-input MAAP set after 2ns using both perfect 20 nm devices and variable-size devices, respectively. The actual results (black squares) are plotted alongside the ideal result (red line). (c)-(d) Histogram of the errors from (a)-(b) respectively.}
\label{fig:VariationHist}
\end{figure}

\begin{figure}
\centering
\includegraphics[scale=0.45]{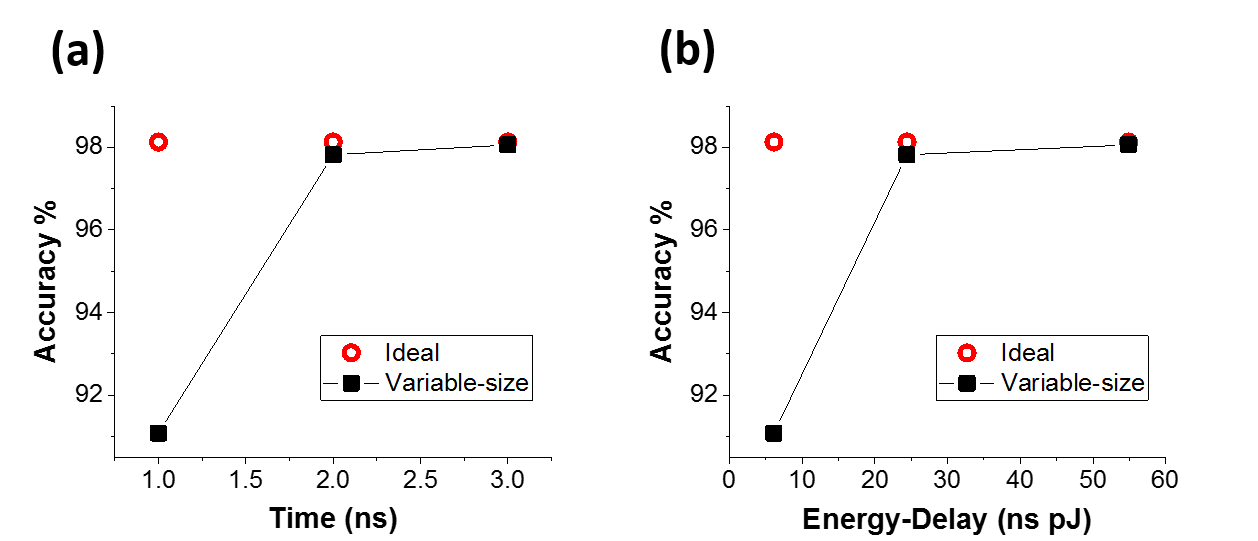}
\caption{(a) Image classification accuracy of MAAP-CNN vs. operational runtime. Results plotted for both ideal 20-nm MTJ side length and variable side length. (b) Classification accuracy vs. energy-delay product.}
\label{fig:VariationED}
\end{figure}

The previous discussion assumed perfect device fabrication with no size variation. Since $I_{c0}$ depends on the FL volume, the bias levels for the MAA sequences will not be precisely accurate if the fabrication is imperfect with small variations in the resulting MTJ size. This results in increased MAAP output error and ultimately lower image classification accuracy. Here we compare the ideal 20 nm-device results to those of 20-nm devices whose edge lengths vary with a deviation of 2.5 nm. In both cases the values of $R_0$ and $I_{c0}$ corresponding to 20 nm devices are used. A comparison of the MAAP set output between the two cases is shown in Fig. \ref{fig:VariationHist}. The device variation results in a distorted output shape in (b) beyond what thermal noise causes in (a). These distorted MAAP outputs cause the image classification accuracy to be more sensitive to MAAP runtime as shown in Fig. \ref{fig:VariationED}. The classification accuracy using variable-size devices is not significantly reduced compared to the ideal devices at 3 ns, but begins to fall off as runtime decreases. These results show that although device variation has a detrimental effect on performance, the effect is by no means crippling. 

\section{Conclusion}
\label{sec:Conclusion}

With the growing interest in MTJs as a framework for neuromorphic computing, new materials and structures are continually being released. This is the time to explore all possible niches for MTJ-neurons to fill in order to take full advantage. With that in mind we proposed this novel neural architecture which harnesses the standard MTJ-neuron to powerful effect, performing multiple layers' worth of CNN computation simultaneously. This saves significant time and energy compared to other CNN implementations that must perform multiple separate computations per layer\cite{CeNNs with CoNNs}. This work also demonstrates remarkable robustness of the MAAP-CNN to thermal noise and even device variation. As the materials, interconnects and crossbar array implementations available continue to improve the potential performance of this architecture will continue to improve.


\begin{thebibliography}{00}

\bibitem{KRoy}
A. Sengupta and K. Roy, ``Encoding Neural and Synaptic Functionalities in Electron Spin: A Pathway to Efficient Neuromorphic Computing,'' Dec. 2017, arXiv:1711.02235v4.

\bibitem{mLogic}
D. Morris, D. Bromberg, J.-G. Zhu and L. Pileggi, ``mLogic: Ultra-Low Voltage Non-Volatile Logic Circuits Using STT-MTJ Devices,'' \textit{Proceedings 49th DAC}, pp. 486--491, Jun. 2012.

\bibitem{Naeemi1} 
{C. Pan and A. Naeemi, ``A Proposal for Energy-Efficient Cellular Neural Network Based on Spintronic Devices,'' \textit{IEEE Trans. Nanotech.,} vol. 15, pp. 820-–827, Sept. 2016, DOI: 10.1109/TNANO.2016.2598147.}

\bibitem{Naeemi2}
{C. Pan and A. Naeemi, ``Non-Boolean Computing Benchmarking for Beyond-CMOS Devices Based on Cellular Neural Network,'' \textit{IEEE J. Expl. Sol.-Stat. Computat. Dev. and Circ.,} vol. 2 pp. 36–-43, Nov. 2016, DOI:10.1109/JXCDC.2016.2633251.}

\bibitem{Spin Switch}
S. Datta, S. Salahuddin and B. Behin-Aein, ``Non-Volatile Spin Switch for Boolean and Non-Boolean Logic,'' \textit{Appl. Phys. Lett.,} 101, pp. 252411--1--5 (2012).

\bibitem{CoNNs}
Y. LeCun, L. Bottou, Y. Bengio and P. Haffner, ``Gradient-Based Learning Applied to Document Recognition,'' \textit{Proc. of the IEEE,} vol. 86, no. 11, pp. 2278--2324, Nov. 1998, DOI:10.1109/5.726791.

\bibitem{Giant TMR}
{H.-X. Liu, Y. Honda, T. Taira, K.-I. Matsuda, M. Arita, T. Uemura and M. Yamamoto, ``Giant Tunneling Magnetoresistance in Epitaxial Co$_2$MnSi/MgO/Co$_2$MnSi Magnetic Tunnel Junctions by Half-Metallicity of Co$_2$MnSi and Coherent Tunneling,'' \textit{Appl. Phys. Lett.,} vol. 101, no. 13, ppl. 132418--1--5, Sep. 2012, DOI:10.1063/1.4755773.}

\bibitem{Crossbar1}
{T. Li, S. Duan, J. Liu and L. Wang, ``An Improved Design of RBF Neural Network Control Algorithm Based on Spintronic Memristor Crossbar Array,'' \textit{Neural Comput. $\&$ Appl.,} vol. 30, no. 6, pp. 1939--1946, Sep. 2018, DOI:10.1007/s00521-016-2715-8.}

\bibitem{Crossbar2}
{Y. Kim, Y. Zhang and P. Li, ``A Reconfigurable Digital Neuromorphic Processor with Memristive Synaptic Crossbar for Cognitive Computing,'' \textit{ACM J. Emerg. Technol. Comput. Syst.,} vol. 11, no. 4, pp. 38:1--38:25, Apr. 2015, DOI:10.1145/2700234.}

\bibitem{Jiaxi}
{J. Hu, G. Stecklein, Y. Anugrah, P. A. Crowell and S. J. Koester, ``Using Programmable Graphene Channels as Weights in Spin-Diffusive Neuromorphic Computing,'' \textit{IEEE JXCDC,} vol. 4, no. 1, pp. 26--34, May 2018, DOI:10.1109/JXCDC.2018.2825299.}

\bibitem{PTM}
{\textit{Predictive Technology Model.} Accessed:Jul. 25, 2017. [Online]. Available: http://ptm.asu.edu}

\bibitem{CeNNs with CoNNs} 
Q. Lou, C. Pan, J. McGuinness, A. Horvath, A. Naeemi, M. Niemier and X. S. Hu, ``A Mixed Signal Architecture for Convolutional Neural Networks,'' Oct. 2018, arXiv:1811.02636v1.

\bibitem{PMA1}
{J. Yun, D. Li, B. Cui, X. Guo, K. Wu, X. Zhang, Y. Wang, Y. Zuo and L. Xi, ``Spin-Orbit Torque Induced Magnetziation Switching in Pt/Co/Ta Structures with Perpendicular Magnetic Anisotropy,'' \textit{J. Phys. D.,} vol. 50, pp. 395001--1--7, Sep. 2017, DOI:10.1088/1361-6463/aa8422.}

\bibitem{PMA2}
{K.-F. Huang, D.-S. Wang, H.-H. Lin and C.-H. Lai, ``Engineering Spin-Orbit Torque in Co/Pt Multilayers with Perpendicular magnetic Anisotropy,'' \textit{Appl. Phys. Lett.} vol. 107, pp. 232407--1--5, Dec. 2015, DOI:10.1063/1.4937443.}

\bibitem{Heusler1}
{J. Zhang, T. Phung, A. Pushp, Y. Ferrante, J. Jeong, C. Rettner, B. P. Hughes, S.-H. Yang, Y. Jiang, and S. S. P. Parkin, ``Bias Dependence of Spin Transfer Torque in Co 2 MnSi Heusler Alloy Based Magnetic Tunnel Junctions,'' \textit{Appl. Phys. Lett.,} vol. 110, no. 17, pp. 172403--1--5, Apr. 2017, DOI:10.1063/1.4981388.}

\bibitem{Heusler2}
{T. Graf, C. Felser, and S. S. P. Parkin, ``Simple Rules for the Understanding of Heusler Compounds,''\textit{Prog. in Sol. Stat. Chem.,} vol. 39, no. 1, pp. 1--50, May 2011, DOI:10.1016/j.progsolidstchem.2011.02.001.}

\bibitem{Heusler3}
{B. Balke, G. H. Fecher, J. Winterlik, and C. Felser, ``Mn$_3$Ga, a Compensated Ferrimagnet with High Curie Temperature and Low Magnetic Moment for Spin Torque Transfer Applications,'' \textit{Appl. Phys. Lett.,} vol. 90, no. 15, pp. 152504--1--3, Apr. 2007, DOI:10.1063/1.2722206.}

\bibitem{Tungsten}
C.-F. Pai, L. Liu, Y. Li, H. W. Tseng, D. C. Ralph and R. A. Buhrman, ``Spin Transfer Torque Devices Utilizing the Giant Spin Hall Effect of Tungsten,'' \textit{Appl. Phys. Lett.,} 101, 122404--1--4 (2012).

\bibitem{Tantalum}
L. Liu, C.-F. Pai, Y. Li, H. W. Tseng, D. C. Ralph and R. A. Buhrman, ``Spin-Torque Switching with the Giant Spin Hall Effect of Tantalum,'' \textit{Science,} vol. 336, pp. 555--558 (2012).

\bibitem{Ralph and Stiles}
{D. C. Ralph and M. D. Stiles, ``Spin Transfer Torques,'' \textit{J. Magn. Magn. Mater.,} vol. 320, pp. 1190-1216, Apr. 2008, DOI:10.1016/j.jmmm.2007.12.019.}

\bibitem{Ispin}
{W.H. Butler, T. Mewes, C. K. A. Mewes, P. B. Visscher, W. H. Rippard, S. E. Russek and R. Heindl, ``Switching Distriutions for Perpendicular Spin-Torque Devices Within the Macrospin Approximation,'' \textit{IEEE TMAG.,} vol. 48, no. 12, pp. 4684--4700, Dec. 2012, DOI:10.1109/TMAG.2012.2209122.}

\end{thebibliography}
\end{document}